\documentclass{svproc}
\usepackage{url}
\usepackage{enumitem}
\usepackage{mathtools}
\usepackage{amsfonts}
\usepackage{amssymb}
\usepackage{amsmath}
\usepackage[export]{adjustbox}
\usepackage{multirow}

\def\etal{\emph{et al.}}
\begin{document}
\mainmatter

\title{Spatially Variant Laplacian Pyramids for Multi-Frame Exposure Fusion}

\titlerunning{Spatially Variant Laplacian Pyramid}

\author{Anmol Biswas\inst{1} \and Green Rosh K S\inst{2} \and
Sachin Deepak Lomte\inst{3}}

\authorrunning{Anmol Biswas et al.}

\tocauthor{Anmol Biswas, Green Rosh K S and Sachine Deepak Lomte}

\institute{Samsung R\&D Institute, Bangalore, \\
\email{anmolbiswas@gmail.com}
\and
Samsung R\&D Institute Bangalore, \\
\email{greerosh.ks@samsung.com}
\and
Samsung R\&D Institute Bangalore, \\
\email{sachin.lomte@samsung.com}
}

\maketitle

\begin{abstract}
Laplacian Pyramid Blending is a commonly used method for several seamless image blending tasks. While the method works well for images with comparable intensity levels, it is often unable to produce artifact free images for applications which handle images with large intensity variation such as exposure fusion. This paper proposes a spatially varying Laplacian Pyramid Blending to blend images with large intensity differences. The proposed method dynamically alters the blending levels during the final stage of Pyramid Reconstruction based on the amount of local intensity variation. The proposed algorithm out performs state-of-the-art methods for image blending both qualitatively as well as quantitatively on publicly available High Dynamic Range (HDR) imaging dataset. Qualitative improvements are demonstrated in terms of details, halos and dark halos. For quantitative comparison, the no-reference perceptual metric MEF-SSIM was used.
\end{abstract}

\section{Introduction}
\label{sec:intro}

\begin{figure}[t]
\begin{center}
    \includegraphics[width=1.0\linewidth]{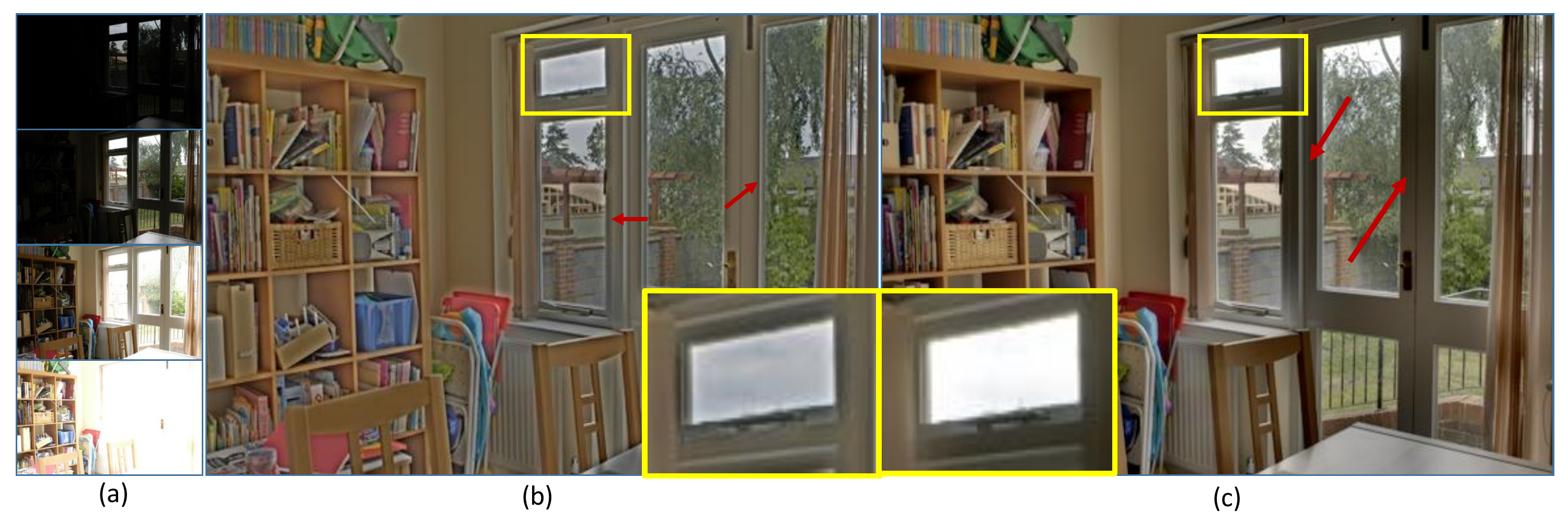}

\end{center}
\vspace{-0.6cm}
\caption{Challenges in Laplacian Pyramid Blending. (a) input exposure stack (b) Blending with 3 levels (c) Blending with 6 levels. With less number of levels(b), the output image looks cartoonish and produces strong gradient reversal artifacts (red arrows). With large number of levels, the image looks more natural, but produces wide halos (red arrows) and less details (yellow inset)}
\label{fig:lapArtifacts}
\end{figure}

Seamlessly blending images together is an important low-level image processing operation that is used in a variety of applications such as, panorama generation, High Dynamic Range (HDR) imaging and so on. Laplacian Pyramid Blending ~\cite{ref:lapBlend} is a popular method for this as it can generate natural looking images at relatively low computational effort. Of these applications, HDR imaging is a method for faithfully capturing the large dynamic range present in a natural scene by a camera with limited sensor capabilities - that can only produce Low Dynamic Range (LDR) images. Typically it is done by exposure fusion - that is, to take multiple LDRs with bracketed exposure times and blend those images. This necessitates blending images with large variations in intensity. There has been a lot of research done to work out the optimal weighting function to weigh the different exposures ~\cite{ref:mertens}~\cite{ref:debevec}~\cite{ref:robertson}~\cite{ref:hdrStar}. Once the weight maps are computed, HDR pipelines typically use Laplacian Pyramid Blending with a certain number of levels to blend these images~\cite{ref:mertens}. From Fig. \ref{fig:lapArtifacts}, it can be observed that keeping a constant number of pyramid levels for blending the entire image forces us to accept some tradeoffs. Lower number of levels produces higher dynamic range, but the blended image has sharp boundary halos and looks unnatural. Higher number of levels creates more natural images, but at the cost of dynamic range, details and spread-out halos.

Several works have tried to address the issue of halos in multi-expsore fusion. Shen \etal~\cite{ref:boostedLap} develops a method where details extracted in different levels are boosted to reduce artifacts using a custom developed weightmap. This method, however is very slow to be implemented in real-time. Li \etal~\cite{ref:wgif} introduces weighted guided image filtering which can be applied for multi-expsosure fusion application. All these methods, however does not take into account spatial structure of the images while deciding amount of detail extraction required; i.e, they use constant number of image pyramid levels throughout the image.

This paper proposes a blending algorithm based on Pyramid Blending which can effectively alter the level of blending spatially in a patch-based manner. This dynamic nature of the blending is dictated by the spatial characteristics of the weight function and works to maximize dynamic range, and minimize halos while otherwise maintaining the naturality of the blended image that comes with using a reasonably large number of pyramid levels for blending

The rest of the paper is organized as follows. Section 2 discusses the proposed method to blend two images taken with different exposures and the pipeline to blend multiple images into one HDR output. Section 3 provides qualitative and quantitative comparisons of the blending method with several state of the art exposure fusion based methods.

\begin{figure}[ht]
\begin{center}
    \includegraphics[width=1.0\linewidth]{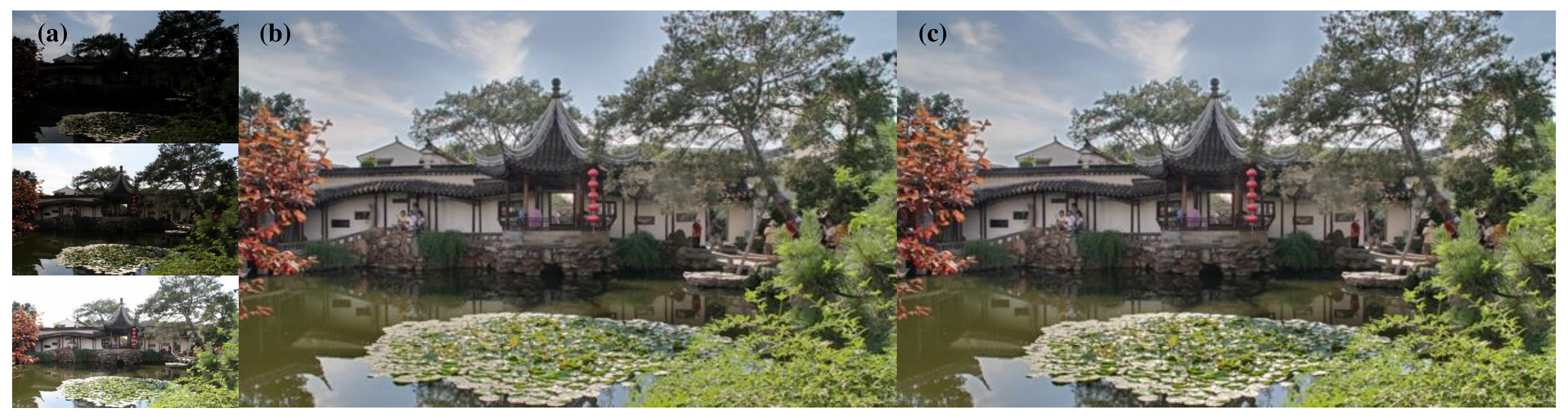}

\end{center}
\vspace{-0.6cm}
\caption{Comparison of average laplacian weights and patch variance weights. (a) Input exposure stack (b) Output of the proposed method with average laplacian weighting (c) Output of the proposed method with patch variance weighting. Weighting by patch variance shows slight improvement in terms of halos, visible around the tree.}
\label{fig:lap_var_comp}
\end{figure}

\begin{figure}[ht]
\begin{center}
    \includegraphics[width=1.0\linewidth]{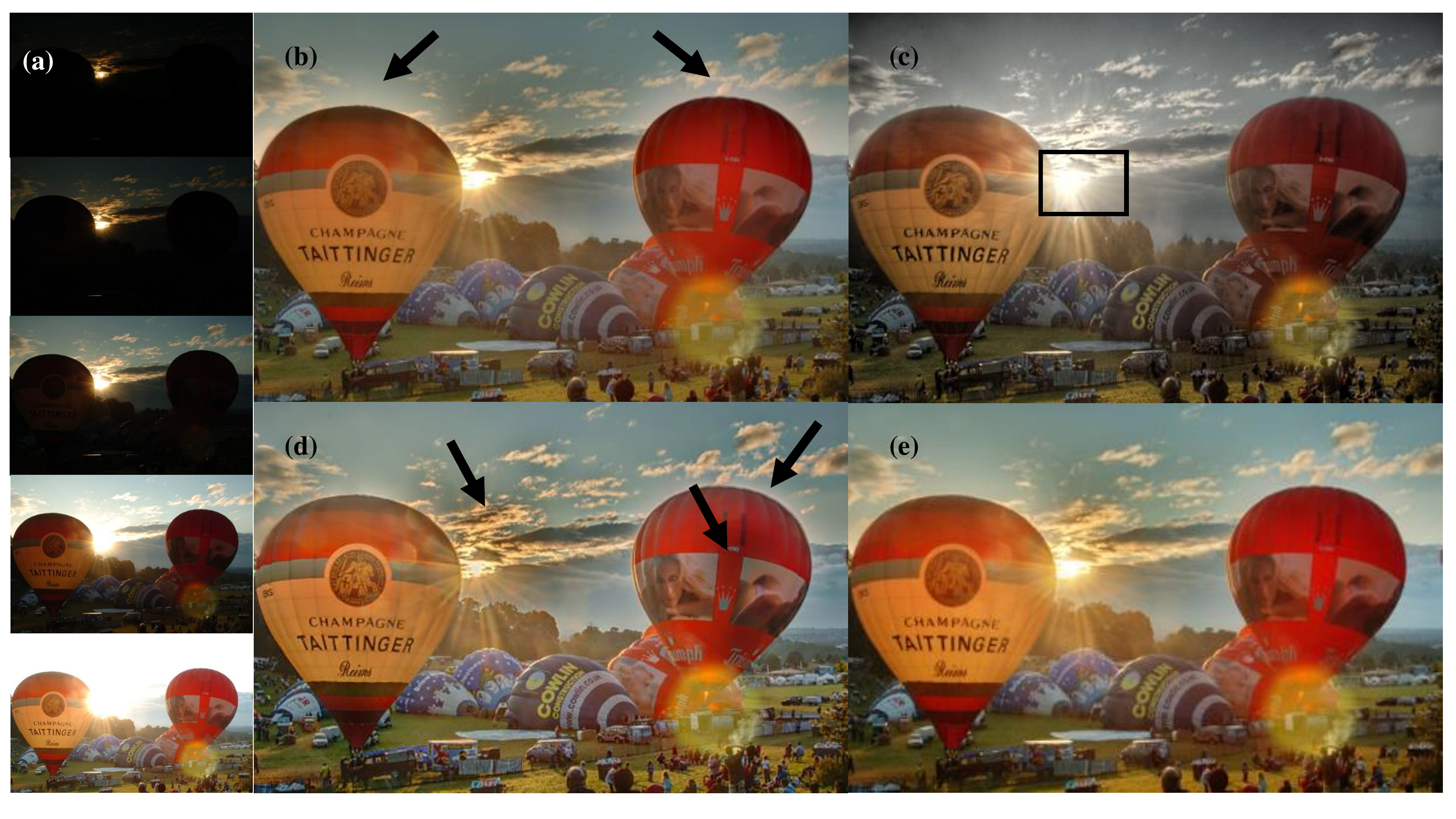}

\end{center}
\vspace{-0.6cm}
\caption{Halos and Gradient Reversal (Dark Patches/Shadows). (a) Input image stack (b) Output of standard Laplacian Pyramid Blending (c) Output of Gu \etal ~\cite{ref:gu12} (d) Output of Boosting Laplacian Pyramid ~\cite{ref:boostedLap} (e) Proposed method with Patch Variance weighting function. Standard Pyramid Blending (b) has Halos and Gradient Reversal (dark patches), Boosting Laplacian Pyramid (d) mainly has dark patches and some halos (black arrows) (c)loses some dynamic range due to saturation around the sun (black box), while (e) has neither halos nor dark patches}
\label{fig:halos1}
\end{figure}

\subsection{Overview of Laplacian Pyramid Blending}

Given two images $I_s$ and $I_h$ and a weight map $W$, laplacian pyramid blending ~\cite{ref:lapBlend} provides a methodology to smoothly blend the images. It consists of three major steps: pyramid decomposition, pyramid blending and pyramid reconstruction. In the pyramid decomposition step, first gaussian pyramids of the images are generated as follows:

\begin{equation}
\label{eq:gaussPyr}
I_{Gauss}^l = (\mathcal{G} \circledast I_{Gauss}^{l-1})\downarrow
\end{equation}

Here, $I_{Gauss}^l$ refers to the $l^{th}$ level of the gaussian pyramid for image $I \in [I_s, I_h, W]$, $\mathcal{G}$ refers to a guassian filter, $\circledast$ refers to the convolution operator and $\downarrow$ denotes a downsampling operator. Next the laplacian pyramids are constructed as follows:

\begin{equation}
I_{Laplacian}^l = 
\begin{cases}
  I_{Gauss}^l - \mathcal{G} \circledast (I_{Gauss}^{l+1}\uparrow) , & \text{if } l < M \\
  I_{Gauss}^l                                                    , & \text{if } l = M
\end{cases}
\end{equation}

Here $I_{Laplacian}^{l}$ refers to the $l^{th}$ level of the laplacian pyramid for image $I \in [I_s, I_h]$, $\uparrow$ refers to an upsampling operator and $M$ refers to the total number of levels in the pyramid.

This is followed by the pyramid blending stage where the Laplacian pyramids are blended to form a blended pyramid $I_{LapBlend}$ as follows:

\begin{equation}
I_{LapBlend}^l = I_{s_{Laplacian}}^l \otimes W_{Gauss}^l + I_{h_{Laplacian}}^l \otimes (1 - W_{Gauss}^l)
\end{equation}.

Here $\otimes$ denotes an element-wise multiplication operator. This is followed by the final pyramid reconstruction stage where each level ($I_{Recon}^{l}$) is reconstructed as follows:

\begin{equation}
I_{Recon}^l = 
\begin{cases}
 \mathcal{G} \circledast (I_{Recon}^{l+1}\uparrow) + I_{LapBlend}^l, & \text{if } l < M \\
 I_{LapBlend}^l                                    , & \text{if } l = M
\end{cases}
\end{equation}

The top-most level of the reconstructed images, i.e, $I_{Recon}^1$ gives the final blended output. While laplacian pyramid blending works well for most image fusion tasks, it produces artifacts for multi-exposure fusion, in regions of large intensity differences. When the total number of levels is large, the algorithm tends to produce wide halos around edges, and when the number of levels is small, strong gradient reversal and halo artifacts are observed (Fig. \ref{fig:lapArtifacts}. To counter these issues, a modified laplacian pyramid blending with spatially varying levels is proposed as described below.

\begin{figure}[ht]
\begin{center}
    \includegraphics[width=1.0\linewidth]{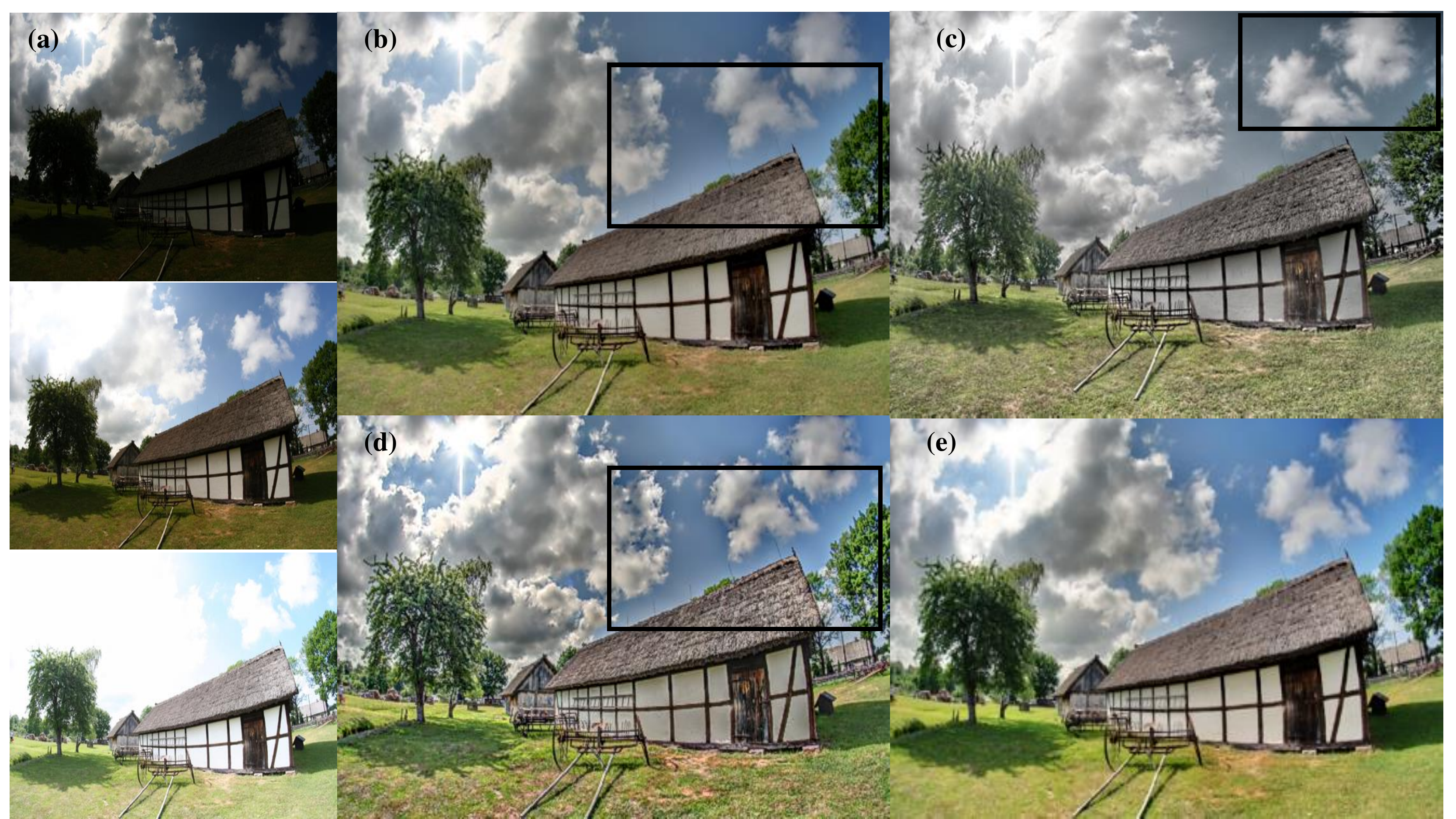}

\end{center}
\vspace{-0.6cm}
\caption{Halos and Gradient Reversal 2 (Dark Patches/Shadows). (a) Input image stack (b) output of standard Laplacian Pyramid Blending (c) Output of Gu \etal ~\cite{ref:gu12} (d) Output of Boosting Laplacian Pyramid  ~\cite{ref:boostedLap}(e) Proposed method with Patch Variance weighting function. Halos in (b) and strong Gradient Reversal in the form of dark shadows around the clouds in (d) are visible inside the black rectangle. (c) shows decoloration and darkening of the sky color (black box)}
\label{fig:halos2}
\end{figure}

\begin{figure}[ht]
\begin{center}
    \includegraphics[width=1.0\linewidth]{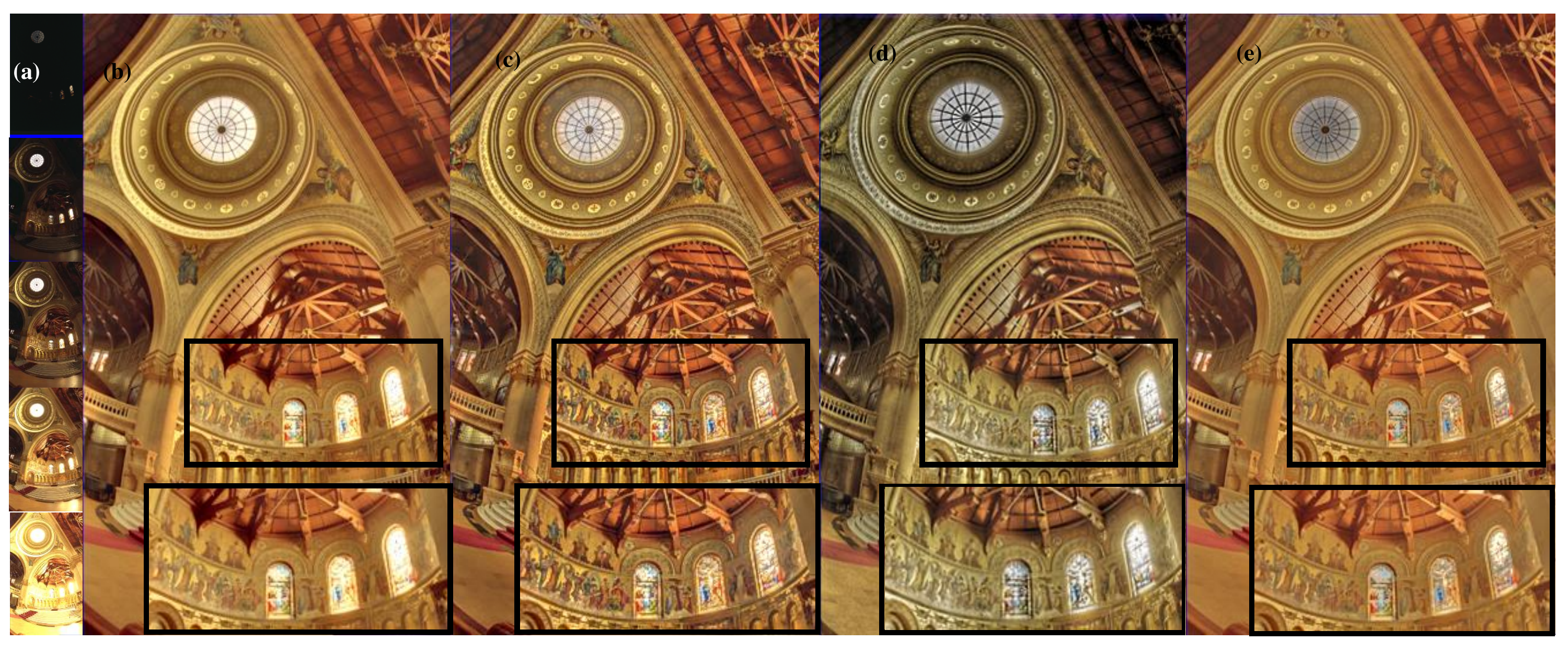}

\end{center}
\vspace{-0.6cm}
\caption{Dynamic Range and Preserving Details. (a) Input image stack (b) output of standard Laplacian Pyramid Blending (c) Output of Boosting Laplacian Pyramid  ~\cite{ref:boostedLap}(d)  Output of Gu \etal ~\cite{ref:gu12} (e) Proposed method with Patch Variance weighting function. Black Inset shows the region where details in saturated region are improved by Boosting Laplacian Pyramid and the proposed method}
\label{fig:details}
\end{figure}

\begin{figure}[ht]
\begin{center}
    \includegraphics[width=1.0\linewidth]{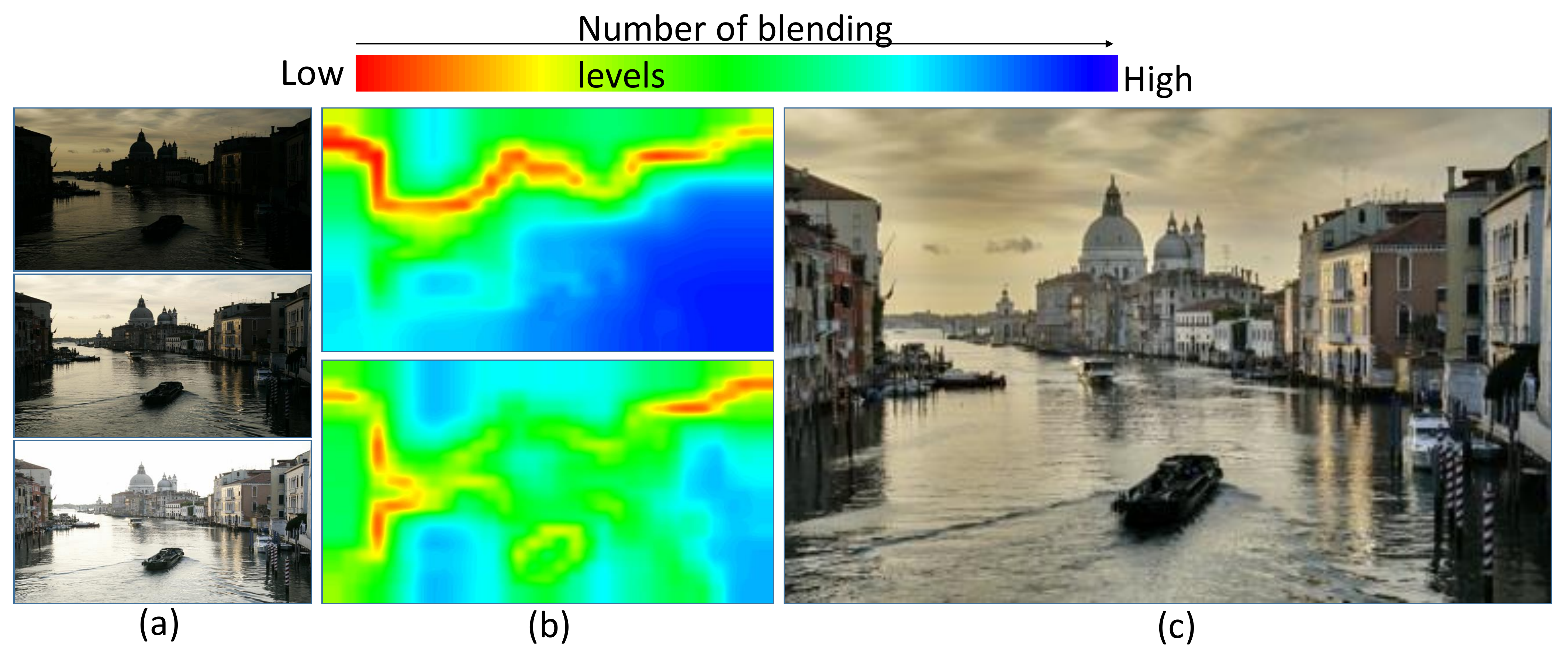}

\end{center}
\vspace{-0.6cm}
\caption{Visualization of spatially varying levels. (a) Input image stack (b) Maps, showing the spatial variation of the equivalent number of levels in the proposed method. Top - Map for blending first two images, Bottom - Map for blending the previous output and third image of the set (c)  Output of the proposed method}
\label{fig:levelMap}
\end{figure}
\section{Proposed Method}
\label{sec:method}

Given a set of $n$ short exposure images $I_s^{1-n}$, high exposure images $I_h^{1-n}$, and $n-1$ weightmaps $W^{1 - (n-1)}$, the aim is to produce a blended HDR image which looks natural and has fewer halos. This paper proposes a method to blend images using spatially varying Laplacian Pyramids to reduce halos and gradient reversal artifacts (dark patches around bright regions). First, a brief overview of the standard Laplacian Pyramid blending ~\cite{ref:lapBlend} is given, followed by the proposed spatially variant Lapalacian Pyramid Blend. For brevity, image blending of two images is detailed in this section. The method can be extended easily to handle multiple images also as shown in section \ref{sec:multiBlend}.

\subsection{Proposed Spatially Variant Level-based Blending} 
Conventional laplacian pyramid blending assumes a uniform number of levels throughout the image. A higher number of levels is essential for a smooth blending. However, this cause halo and gradient reversal artifacts in regions of large intensity changes in input images. Hence a pyramid blending algorithm which takes into the account the intensity variations in the input image, and dynamically changes the level of blending applied at each region is developed. The proposed method also processes the inputs in three steps similar to laplacian pyramid blend ~\cite{ref:lapBlend}. The first stage of pyramid decomposition is identical to the pyramid decomposition step in ~\cite{ref:lapBlend}. In the second stage of pyramid blending, an additional Gaussian Blend pyramid ($I_{GaussBlend}$) is also constructed along with $I_{LapBlend}$ as follows:

\begin{equation}
I_{GaussBlend}^l = I_{s_{Gauss}}^l \otimes W_{Gauss}^l + I_{h_{Gauss}}^l \otimes (1 - W_{Gauss}^l)
\end{equation}

To bring in the intensity variation, the images are processed in overlapping patches of $K \times K$ at each level during the third stage of pyramid reconstruction. At each level of pyramid reconstruction, each of the $K \times K$ are reconstructed as follows:

\begin{equation}
I_{p_{Recon}}^l = 
\begin{cases}
 (1 - \alpha).(\mathcal{G} \circledast (I_{p_{Recon}}^{l+1}\uparrow) + I_{p_{LapBlend}}^l) + \alpha.I_{p_{GaussBlend}}^l, & \text{if } l < M \\
 I_{p_{LapBlend}}^l                                                                             , & \text{if } l = M
\end{cases}
\end{equation}

Here $I_p$ refers to a $K \times K$ patch extracted from a given layer from Blended Laplacian or Gaussian Pyramids or Reconstructed pyramid. Here i$\alpha$ refers to a dynamically computed weighing factor which encompasses the intensity variations for a given patch. A higher value of $\alpha$ denotes a larger variation in intensity, and correspondingly a lower level of blending. To understand this, consider an $alpha$ value of 1 . In this case the the reconstructed image at a level $l$ is same as the Gaussian blended image. This effectively cancels out any contribution due to higher number of levels. 

The weighing factor $\alpha$ is modelled as a function of the input image, $\alpha = f(\mathcal{M}(I_h)$. Here $\mathcal{M}$ is a map derived from image $I_h$ as follows:

\begin{equation}
\mathcal{M}(I) = \exp(-\frac{(I - 1.0)^2}{2\sigma^2})
\end{equation}

Here $I$ is normalized to the range $[0,1]$ and the value of $\sigma$ is empirically chosen to be $0.3$. 

The choice of the function $f()$ should be representative of the intensity changes in the given patch. Two different functions for $f()$ were experimented with, as described below.

\subsubsection{Laplacian of Patch.}

The laplacian of any image patch gives the gradient information present in the patch, which is indicative of the amount of intensity variation in the patch. Hence for the first choice of the weighing function, the mean of the intensities of the laplacian of the patches extracted from $\mathcal{M}(I_h)$ was taken.

\subsubsection{Variance of Patch.}

The variance of the image pixel intensities also give a good representation of the spatial variation of intensity. Hence the variance of the patches extracted from $\mathcal{M}(I_h)$ as $f()$ was also experimented with. The outputs of the two proposed methods are compared in section \ref{sec:expt}.

\subsubsection{Blending of patches}

Once the patches for each level are reconstructed they have to be blended in a seamless manner to generate the full image. Modified raised cosine filter proposed by ~\cite{ref:cosineFilter} was used to blend the overlapping patches. It is to be noted that each of the patches have an overlapping factor of $50\%$.

\subsection{Extension to multiple frames}
\label{sec:multiBlend}
Given $N$ differently exposed images and $N-1$ weight maps, the proposed two-frame blending method can be sequentially executed $N-1$ times to generate the output as follows:

\begin{equation}
I_o^i = \mathcal{B}(I_o^{i-1}, I_i, W_i)
\end{equation}

Here $I_o^i$ represents HDR output formed by blending $i$ frames and $I_i$ denotes the $i^{th}$ input image. $\mathcal{B}$ refers to the proposed blending algorithm.

\section{Experimental Results}

\label{sec:expt}
\subsection{Qualitative Comparisons}

The proposed exposure fusion method was evaluated on publicly available HDR dataset ~\cite{ref:mefssim}. Fig. \ref{fig:lap_var_comp} compares the two variants of the proposed method. It is observed that patch variance as the weight factor produces slightly superior outputs. Henceforth, patch variance is used for all comparisons. The output of the proposed method with patch variance for weighting factor are compared with the output from using standard Laplacian Pyramid Blending with the same number of levels and identical weight map and the outputs of Boosting laplacian Pyramid ~\cite{ref:boostedLap} and Gu \etal ~\cite{ref:gu12} in Fig. \ref{fig:halos1}, Fig. \ref{fig:halos2} and Fig. \ref{fig:details}. Halos, dark patches arising from gradient reversal and level of details are specifically compared. It is observed that the proposed method is the most effective at eliminating halos and dark patches while maintaining a level of details and dynamic range roughly similar to Boosting Laplacian Pyramid.

\begin{table}[h]
\begin{center}
\begin{tabular}{|c|c|c|c|c|}
\hline
\bf{Image Set} & \bf{Gu \etal~\cite{ref:gu12}} & \bf{Raman \etal~\cite{ref:raman09}} & \bf{Shen \etal~\cite{ref:boostedLap}} & \bf{Proposed}  \\
\hline
\hline
Balloons & 0.913 & 0.768 & 0.902 & \bf{0.936}\\
\hline
Belgium House  & 0.896 & 0.809 & 0.915 & \bf{0.933} \\
\hline
Chinese Garden & 0.927 & 0.911 & 0.917 & \bf{0.96}\\
\hline
Kluki & \bf{0.921} & 0.901 & 0.861 & 0.915 \\
\hline
Memorial & 0.87 & 0.617 & 0.909 & \bf{0.94} \\
\hline
Cave	& \bf{0.933} & 0.693 & 0.922 & 0.914 \\
\hline
Farmhouse & 0.932 & 0.877  & 0.942 & \bf{0.971} \\
\hline
House & \bf{0.876} & 0.77 & 0.876 & 0.829 \\
\hline
Lamp & 0.871 & 0.864 & 0.875 & \bf{0.902} \\
\hline
Landscape & 0.94 & 0.953 & 0.872 & \bf{0.954} \\
\hline
LightHouse & 0.934 & 0.938 & 0.873 & \bf{0.962} \\
\hline
Office & 0.899 & 0.906 & 0.93 & \bf{0.962} \\
\hline
Tower & 0.931 & 0.895 & 0.873 & \bf{0.965} \\
\hline
Venice & 0.889 & 0.892 & 0.868 & \bf{0.915} \\
\hline
\hline
\bf{Mean} & 0.909 & 0.842 & 0.895 & \bf{0.932} \\
\hline
\end{tabular}
\vspace{0.2cm}
\caption{Comparisons using MEF-SSIM}
\label{tab:mefssim}
\end{center}
\end{table}
\subsection{Quantitative Comparisons}

Quantitative comparison results against the boosted Laplacian Pyramid proposed by Shen \etal ~\cite{ref:boostedLap}, gradient field exposure fusion proposed by Gu \etal ~\cite{ref:gu12} and bilateral filter based multi-exposure compositing, proposed by Raman \etal ~\cite{ref:raman09}  are provided. The metric MEF-SSIM ~\cite{ref:mefssim} proposed by Ma \etal ~\cite{ref:mefssim} is being used for comparison. This metric is chosen for the objective evaluation of the proposed algorithm since it provides a way to analyse the perceptual image quality without the need of any reference images. The metric is based on the multi-scale SSIM principle ~\cite{ref:ssim} and measures the local structure preservation of the output image with respect to input images at fine scales and luminance consistency at coarser scales. For a more detailed discussion on the metric, the readers may refer to ~\cite{ref:mefssim}.

The algorithms are evaluated on the public dataset provided by ~\cite{ref:mefssim}. The results are summarized in Table \ref{tab:mefssim}.

From the table, it can be observed that the proposed method produces much better scores in most of the scenes, and produces a better overall score. This shows that the algorithm can generalize very well compared to the state-of-the-art method.

\subsection{Visualizations}
Fig. \ref{fig:levelMap} shows the spatial variation of the equivalent number of blending levels in the proposed method. It clearly indicates how the level of blending is reduced in and around regions with high intensity variation and how the effect slowly dies out with distance. This variation allows the algorithm to control halos while also maintaining the naturalness of the final output image.

\section{Conclusion.}

This paper proposes a novel method for blending images with large variation in intensity. It overcomes the limitations of Laplacian Pyramid Blending by dynamically altering the blending level on a spatially varying basis. This allows us to leverage a range of blending levels in a single image as opposed to conventional methods. Extensive qualitative comparisons with state of the art methods in exposure fusion shows the ability of the proposed algorithm to  generate artifact free images with greater dynamic range and details. A quantitative study using MEF SSIM scores is also performed to show that the method produces perceptually superior results.

%
%
%
%

\bibliographystyle{splncs}
\bibliography{146.source}

\begin{thebibliography}{10}

\bibitem{ref:lapBlend}
Burt, P., Adelson, E.:
\newblock The laplacian pyramid as a compact image code.
\newblock IEEE Transactions on communications \textbf{31} (1983)  532--540

\bibitem{ref:mertens}
Mertens, T., Kautz, J., Van~Reeth, F.:
\newblock Exposure fusion: A simple and practical alternative to high dynamic
  range photography.
\newblock In: Computer graphics forum. Volume~28., Wiley Online Library (2009)
  161--171

\bibitem{ref:debevec}
Debevec, P.E., Malik, J.:
\newblock Recovering high dynamic range radiance maps from photographs.
\newblock In: ACM SIGGRAPH 2008 classes, ACM (2008) ~31

\bibitem{ref:robertson}
Robertson, M.A., Borman, S., Stevenson, R.L.:
\newblock Dynamic range improvement through multiple exposures.
\newblock In: Proceedings 1999 International Conference on Image Processing
  (Cat. 99CH36348). Volume~3., IEEE (1999)  159--163

\bibitem{ref:hdrStar}
Tursun, O.T., Aky{\"u}z, A.O., Erdem, A., Erdem, E.:
\newblock The state of the art in hdr deghosting: a survey and evaluation.
\newblock In: Computer Graphics Forum. Volume~34., Wiley Online Library (2015)
  683--707

\bibitem{ref:boostedLap}
Shen, J., Zhao, Y., Yan, S., Li, X.,  et~al.:
\newblock Exposure fusion using boosting laplacian pyramid.
\newblock IEEE Trans. Cybernetics \textbf{44} (2014)  1579--1590

\bibitem{ref:wgif}
Li, Z., Zheng, J., Zhu, Z., Yao, W., Wu, S.:
\newblock Weighted guided image filtering.
\newblock IEEE Transactions on Image processing \textbf{24} (2014)  120--129

\bibitem{ref:gu12}
Gu, B., Li, W., Wong, J., Zhu, M., Wang, M.:
\newblock Gradient field multi-exposure images fusion for high dynamic range
  image visualization.
\newblock Journal of Visual Communication and Image Representation \textbf{23}
  (2012)  604--610

\bibitem{ref:cosineFilter}
Hasinoff, S.W., Sharlet, D., Geiss, R., Adams, A., Barron, J.T., Kainz, F.,
  Chen, J., Levoy, M.:
\newblock Burst photography for high dynamic range and low-light imaging on
  mobile cameras.
\newblock ACM Transactions on Graphics (TOG) \textbf{35} (2016)  192

\bibitem{ref:mefssim}
Ma, K., Zeng, K., Wang, Z.:
\newblock Perceptual quality assessment for multi-exposure image fusion.
\newblock IEEE Transactions on Image Processing \textbf{24} (2015)  3345--3356

\bibitem{ref:raman09}
Raman, S., Chaudhuri, S.:
\newblock Bilateral filter based compositing for variable exposure photography.
\newblock In: Eurographics (short papers). (2009)  1--4

\bibitem{ref:ssim}
Wang, Z., Bovik, A.C., Sheikh, H.R., Simoncelli, E.P.,  et~al.:
\newblock Image quality assessment: from error visibility to structural
  similarity.
\newblock IEEE transactions on image processing \textbf{13} (2004)  600--612

\end{thebibliography}
\end{document}